\documentstyle[preprint,aps]{revtex}
\input{psfig}
\begin{document}
\pagestyle{myheadings}
\markright{Submitted to Phys. Fluids}
\draft
\title{An Intermittency Model For Passive-Scalar Turbulence}
\author{Nianzheng Cao and Shiyi Chen}
\address{IBM Research Division, T. J. Watson Research Center,
P.O. Box 218, \\Yorktown Heights, NY 10598 and\\
Theoretical Division and Center for Nonlinear Studies,
Los Alamos National Laboratory,\\ Los Alamos, NM 87545}
\maketitle

\begin{abstract}

A phenomenological model for the inertial range scaling of passive-scalar 
turbulence is developed based on a bivariate log-Poisson model.  
An analytical formula of the scaling exponent for three-dimensional 
passive-scalar turbulence is deduced. The predicted scaling exponents 
are compared with experimental measurements, showing good agreement.

\end{abstract}


The inertial range dynamics of a passive scalar 
advected by fluid turbulence is of great theoretical 
interest\cite{andy}. In particular, recent success\cite{k94} in the 
deduction of the anomalous scaling exponent for the passive scalar field 
advected by a white and Gaussian velocity has inspired renewed 
enthusiasm for the problem\cite{papers}.

For the three dimensional turbulence diffusion, 
the advective velocity field is governed by the
Navier-Stokes equation whose statistics is far from white and
Gaussian. The extension of Kraichnan's theory\cite{k94} for this problem
is intriguing and non-trivial. In fact it has even been a 
difficult task to experimentally 
measure the scaling exponents of the passive scalar. 
The existing experimental data do not yield
convergent results\cite{sreeni,meneveau,antonia}.
Direct numerical simulation (DNS) has been
a useful alternative for studying scaling behaviors in 
fluid turbulence\cite{cao}. In a recent note, using the DNS data, 
we have tried to explain the discrepancies among experimental
results\cite{cao1}. 

The existing phenomenological theories, such as the $\beta$ 
model\cite{beta} and a bivariate log-Normal model\cite{van}, are based on
fractal structures of turbulence\cite{meneveau}, 
yielding analytical predictions
of scaling exponents. Nevertheless, it has been noted 
that log-normal distribution leads to 
negative values of scaling exponents for high
order passive scalar structure functions, contradicting existing 
numerical and experimental measurements. 

In this letter, we develop a phenomenological theory
based on a bivariate log-Poisson model. The
analytical formula for the scaling exponents is obtained.
The scaling exponents from the theory are compared with
those from experimental measurements, showing good agreement. 

According to Kolmogorov's refined similarity hypothesis (RSH), 
or K62 theory\cite{k62} for the velocity field and similar RSH 
for the passive scalar field proposed by Obukhov\cite{obu}, 
\begin{eqnarray}
\Delta u_\ell=v_u(\ell\epsilon_{\ell})^{1/3}, \nonumber \\
\Delta T_\ell = v_{T} \ell^{1/3}\epsilon_{\ell}^{-1/6}N_{\ell}^{1/2}, 
\label{rsh}
\end{eqnarray}
where $\Delta u_\ell = u({\bf x} + \ell) - u({\bf x})$ and
$\Delta T_\ell = T({\bf x} + \ell) - T({\bf x})$ are 
the longitudinal velocity increment and the scalar increment respectively,
$l$ is in the inertial range, 
 $v_u$ and and $v_T$ are random variables 
whose statistics only depend on the Reynolds number. $\epsilon_{\ell}$ 
and $N_{\ell}$ are the locally averaged velocity dissipation 
and scalar dissipation, respectively\cite{sreeni}. 
Equation (\ref{rsh}) has been verified by
experiments and is in general believed to be correct\cite{sreeni1}.
Suppose that the $p-th$ order velocity and passive scalar structure 
functions satisfy the scaling relations in the inertial range: 
$S_p(\ell) = \langle (\Delta u_\ell)^p \rangle \sim \ell^{\zeta_p}$, 
$S_p^{T}(\ell) = \langle (\Delta T_\ell)^p \rangle \sim \ell^{z_p}$,
then to deduce scaling exponents from (\ref{rsh}), a probability 
density function (PDF) of velocity field and
a joint PDF for velocity and passive scalar must be modeled.
Here $\langle \cdot\rangle $ denotes an ensemble and $\zeta_p$ and
$z_p$ are the $p-th$ order scaling exponents for the velocity and 
scalar structure functions, respectively. 
The log-normal assumption in K62 theory\cite{k62} has 
been challenged recently by She and Leveque\cite{she} who argued
that a log-Poisson PDF for $\epsilon_r$ leads to a better description
of the statistics of $\epsilon_r$. A hierarchy model based on the 
log-Poisson PDF was derived for the inertial range scaling exponents, 
showing good agreement with numerical simulations\cite{cao} and
experiments\cite{benzi}. 

The distribution function of a random 
variable $x$ satisfying Poisson distribution can be written as\cite{feller}:
$p(x=i)=p_i=e^a a^i/i!$, where $a$ is the variance 
and the mean for the variable. The generating function of $p_i$ has the form:
$P(s)=\sum_i p_i s^i = e^{-a+as}$. The generating function for a
 bivariate Poisson distribution $p_{i,j}=p(x=i,y=j)$ 
can be defined as 
\[ P(s_1,s_2)=\sum_{ij}p_{ij}s_1^is_2^j=e^{-a_1-a_2-b+a_1s_1+a_2s_2+bs_1s_2},\]
where $a_1$, $a_2$ are variances for $x$ and $y$, respectively, and
$b$ is a constant representing the correlation between $x$ and $y$. 
Let us assume that the velocity dissipation and
passive scalar dissipation have the hierarchy structure relation:
$\epsilon_{\ell_1}=W_{\ell_1\ell_2}\epsilon_{\ell_2}$ and 
$N_{\ell_1}=V_{\ell_1\ell_2}N_{\ell_2}$, 
where $\ell_1$ and $\ell_2$ are two length scales; $W_{\ell_1\ell_2}$
and $V_{\ell_1\ell_2}$ are multiplicative factors depending on $\ell_1$
and $\ell_2$. For $\ell_1$ and $\ell_2$ within the 
inertial range, the multiplicative factors can be written in
the forms of $W_{\ell_1\ell_2}=(\ell_1/\ell_2)^{h}\beta^X$ and
$V_{\ell_1\ell_2}=(\ell_1/\ell_2)^{h_T}\beta_T^Y$, where $\beta$,
$\beta_T$, $h$ and $h_T$ are constants to be determined 
later; $(X,Y)$ are stochastic variables. In the following, we assume 
that $(X,Y)$ follow the bivariate Poisson distribution $p_{i,j}$. 

If there is a scaling, then for $\ell$ in the inertial range, 
$\langle\epsilon_{\ell}^p N_{\ell}^q\rangle\sim\ell^{\tau(p,q)}$, 
where $\tau_{p,q}$ is the scaling
exponent of power order $p$ and $q$. Using the bivariate Poisson distribution
assumption, we have:
\[
\langle W_{\ell_1\ell_2}^pV_{\ell_1\ell_2}^q \rangle= 
(\ell_1/\ell_2)^{ph+qh_T} \sum_{i,j}\beta^{ip}\beta_T^{jq}p_{ij}.
\] 
It is easy to recognize that the summation in the right hand side of the
above equation is nothing but the generating function for the bivariate
Poisson distribution. It is then straightforward to write a formula for
$\tau(p,q)$: 
\begin{equation}
\tau(p,q)= -h p-h_Tq + (a+a_T+b-a\beta^p-a_T\beta_T^q- 
  b\beta^p\beta_T^q)/ln(\ell_1/\ell_2),
\end{equation}
where constants $a$, $a_T$ and $b$ are functions of 
$\ell_1$ and $\ell_2$. The obvious physical constrains 
$\tau(1,0)=\tau(0,1)=0$ result the following relations: 
$a=[h/(1-\beta)]ln(\ell_1/\ell_2-b)$ and
$a_T=[h_T/(1-\beta_T)]ln(\ell_1/\ell_2-b)$. 
If we further assume the correlation between $\epsilon_{\ell}$ and
$N_{\ell}$ satisfying power law scaling in the inertial range, i.e.
$b=\gamma_bln(\ell_1/\ell_2)$ and $\gamma_b$ is a constant, 
we obtain the scaling exponent relation: 
\begin{equation}
\tau(p,q)=\tau_p+\tau_q+\tau_b(p,q),
\end{equation}
where
\[ \tau_p=-h p+[h/(1-\beta)](1-\beta^p), \]
and
\[ \tau_q=-h_Tq+[h_T/(1-\beta_T)](1-\beta_T^q), \]
are scaling exponents for moments $\langle\epsilon_{\ell}^p\rangle$ and
$\langle N_{\ell}^q\rangle$ respectively, and 
\[ \tau_b(p,q)=\gamma_b(1-\beta^p-\beta_T^q+\beta^p\beta_T^q)\]
is the contribution from the cross correlation. There have been several  
discussions about how 
to determine the model constants in $\tau_{p,q}$\cite{she,chen-cao}. 
The parameter $\beta$ depends on the constant $h$ and the intermittency 
parameter $\mu = \tau_2$\cite{chen-cao}, i.e. 
$ \beta=1-\mu/h. $
It has been argued that the 
co-dimension, $C$, for the most singular structure of the velocity
dissipation\cite{she} is related to $\beta$ and $h$:
\begin{equation}
C=h/(1-\beta)=h^2/\mu.
\label{co-d}
\end{equation}

\bigskip
\psfig{file=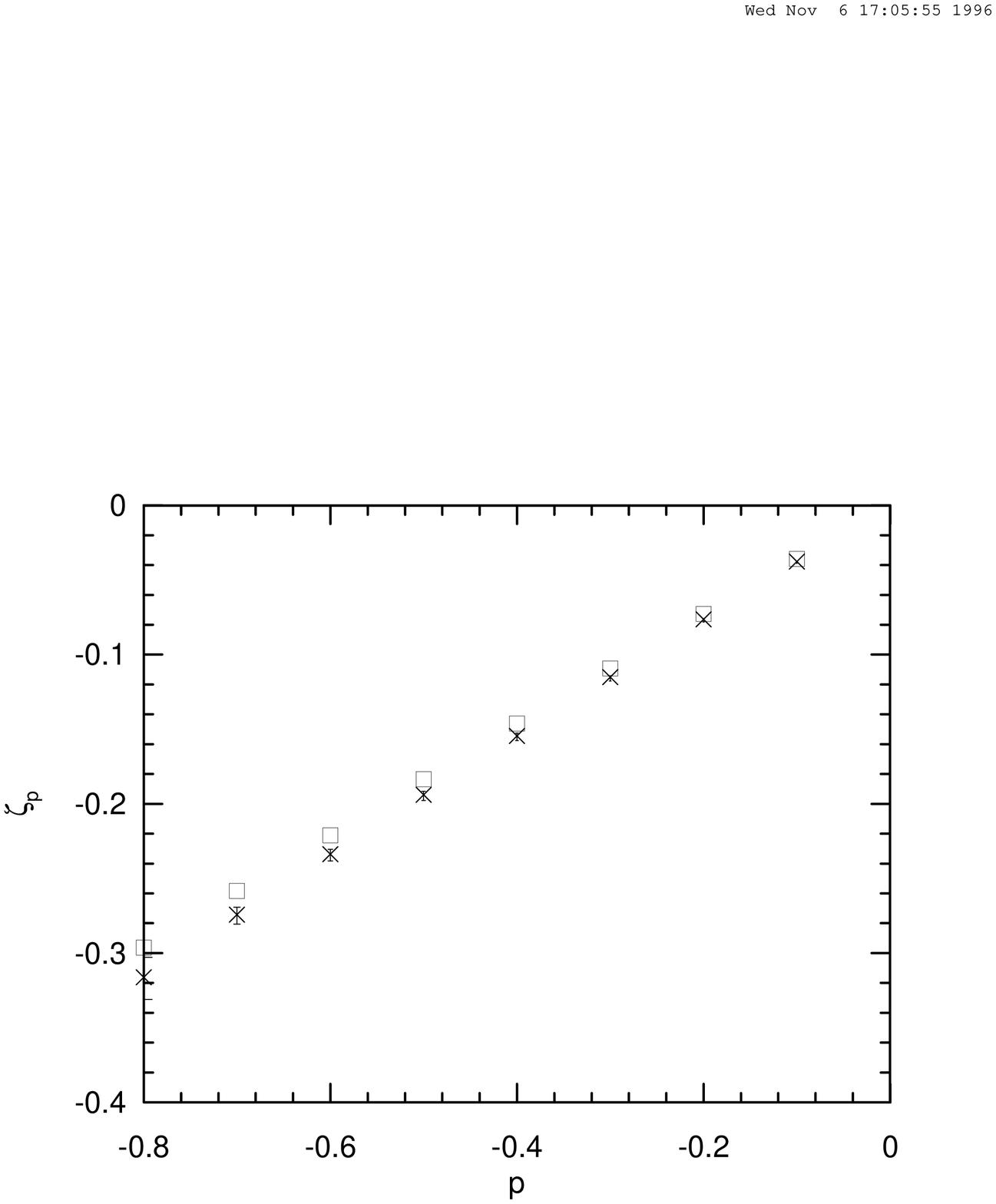,width=220pt}
\noindent
\small {FIG.~1.
Theoretical scaling exponents $\zeta_p$ ($p \le 0$) for
the velocity structure function field as a function
of $p$ ($\times$), compared with numerical measurements 
($\Box$) \cite{cao2}.}
\bigskip

The original hierarchy structure model by She-Leveque
assumes that the high intensity dissipation region tends to
form tube-like structures which implies: $C=2$. Using
$\mu=2/9$\cite{sreeni}, we obtain $h=\pm2/3$. 
The sign associated 
with $h$ is due to the square operation in (\ref{co-d}).
If we further assume $z_p$ grows slower than exponential as $p$ 
tends to $\pm\infty$, the positive (or negative) $h$ 
has to be chosen for the positive (or negative) 
power $p$ in (\ref{co-d}). The original SL model with 
$h=2/3$ corresponds to the range of positive powers, yielding 
$\beta=h=2/3$. As shown in (\ref{rsh}) for $S_p^T$ with $p\ge 0$,
only the negative power $p$ for $\tau_p$ is involved. Hence, for
$p \le 0$, we choose $h=-2/3$, yielding $\beta=4/3$ and: 
\begin{equation}
\tau_p = \frac{2}{3}p + 2 [1 - (4/3)^p].
\label{tau-p}
\end{equation}
Using RSH relation in (\ref{rsh}), we have 
\begin{equation}
\zeta_p = \frac{5}{9}p + 2 [1 - (4/3)^{p/3}]. 
\label{zeta-p}
\end{equation}
In a recent paper\cite{cao2},
we have compared DNS data with the theoretical prediction 
using $h=2/3$ for $\zeta_p$ when $-1 \le p \le 0$. In Fig. 1,
we show again this comparison with (\ref{tau-p}) for
$h=-2/3$.

The similar procedure can be carried out for the passive scalar 
dissipation.  For $q \ge 0$, only the positive $q$ in 
$\tau_q$ has contribution.
Consequently, a positive $h_T$ has to be chosen. 
Using the similar arguments as for the velocity
field, we finally obtain $C_T=h_T/(1-\beta_T)=h_T^2/\mu_T$, 
where $\mu_T$ is the intermittency parameter for
the passive scalar dissipation $N_{\ell}$. 
It is noted that the high intensity region for the
passive scalar dissipation $N$ tends to form sheet-like 
structures\cite{cao1}, implying the co-dimension $C_T = 1$.
We then obtain $h_T=\sqrt{\mu_T}$ and
$\beta_T=1-\sqrt{\mu_T}$. From numerical\cite{cao1} and 
experimental\cite{antonia} data, $\mu_T \approx 1/4$, we have: 
\begin{equation}
\tau_q = -\frac{1}{2}q + [1 - (1/2)^q],
\label{tau-q}
\end{equation}

From (\ref{tau-p}) and (\ref{tau-q}), 
$\tau(p,q)$ can be written as follows:
\begin{equation}
\tau(p,q) =  3 + \frac{2}{3}p -2(\frac{4}{3})^p - \frac{1}{2}q 
-(1/2)^q + \gamma_b [1-(\frac{4}{3})^p - (1/2)^q + (\frac{4}{3})^p(1/2)^q].
\label{tau-pq}
\end{equation}
Using (\ref{rsh}) and (\ref{tau-pq}), we have the following analytical 
formula for $z_p$:
\begin{eqnarray}
z_p= & \frac{1}{3}p+\tau(-\frac{p}{6},\frac{p}{2}) \nonumber \\
   = & 3-\frac{1}{36}p-2(3/4)^{p/6}-(1/2)^{p/2} \nonumber \\
     & +\gamma_b[1-(3/4)^{p/6}-(1/2)^{p/2}+(3/4)^{p/6}(1/2)^{p/2}].
\label{z-p}
\end{eqnarray}
In the above equation, $\gamma_b$ is not determined. 
At this moment, there is no direct 
measurement of $\gamma_b$ and we treat it as a free parameter. 
In Fig.~2a we plot $z_p$ as functions of $p$ for
$\gamma_b = 0, 0.1$ and $0.2$. The scaling exponents from
the experimental measurement carried out by 
Antonia {\em et. al.}\cite{antonia} have also been included. 
A good agreement between theory and measurement for $\gamma_b \approx 0$
is seen, indicating that the correlation between
$\epsilon_{\ell}$ and $N_{\ell}$ is weak. 
In fact, the variation of the scaling exponents is not drastic for 
$\gamma_b$ varying from $0$ to $0.2$. 
In Fig.~2b, we demonstrate the model
dependence on co-dimension $C_T$. We notice that  when $C_T$ is between
0.75 and 1, the model results agree with the experiment.
As $C_T$ increases further, the model prediction deviates from the
experimental data.
It has been found that the co-dimension of passive scalar
field is around 0.6 \cite{sreeni2}. It would be useful to directly 
measure $C_T$, the co-dimension of passive scalar dissipation using 
experimental or numerical simulation data.

The evidence of scaling exponents for
$\gamma_b = 0$ matches well with those from experimental 
measurements might indicate real physics. As a matter of fact,
from (\ref{rsh}) we notice that for positive and large p values,  
the major contribution of $\epsilon_{\ell}^{-p/6}$ to $S_p^T$ comes
from small amplitude events whose structures are blob-like, 
whereas the major contribution of $N_{\ell}^{p/2}$ to $S_p^T$
comes  from large amplitude events whose structures are sheet-like.
The former is often associated with weak stretching regions while 
the latter is associated with strong stretching regions. We suspect that
the weak spatial correlation of these regions 
might be the direct reason why the nearly-independent log-Poisson 
with $\gamma_b \approx 0$ gives a good prediction of the 
scaling exponents. We should mention that (\ref{z-p}) fails to be valid
if power $p$ is too large. In fact, when $p$ is larger than 36 and $\gamma_b$
equals to zero, $z_p$ starts be negative. Up to now, there
is very little understanding about the statistics of very high moments,
due to the difficulties in both experiments or numerical
simulations. Consequently, there is no solid justification for using the 
RSH and Hierarchy model in such high power range.

\bigskip
\psfig{file=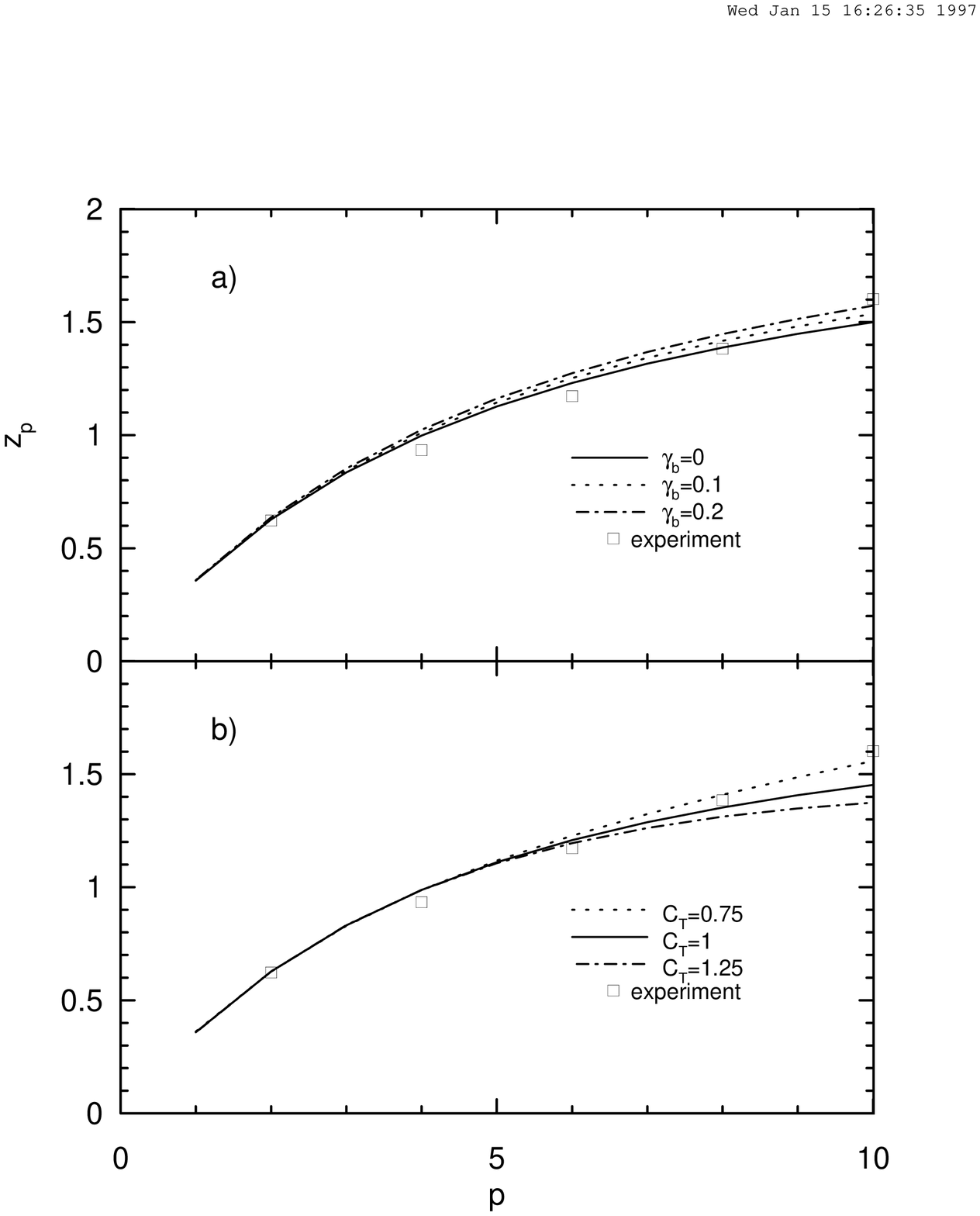,width=220pt}
\noindent
\small {FIG.~2. 
Theoretical scaling exponents $z_p$ for the passive scalar field as functions
of $p$ compared with experimental measurements\cite{antonia}, showing 
the model dependence on: (a) $\gamma_b$ while the $C_T$ equals 1; (b) $C_T$
while $\gamma_b$ equals 0.}





\end{document}